\documentclass[11pt]{article}
\usepackage{amsmath}
\usepackage{amsfonts}
\usepackage{amssymb}
\usepackage{theorem}

\def\Th{\Theta}
\def\p{\partial}
\newtheorem{prop}{Proposition}
\newtheorem{lemma}{Lemma}

\newcommand{\be}{\begin{equation}}
\newcommand{\ee}{\end{equation}}
\newcommand{\bea}{\begin{eqnarray}}
\newcommand{\eea}{\end{eqnarray}}
\newcommand{\beaa}{\begin{eqnarray*}}
\newcommand{\eeaa}{\end{eqnarray*}}

\newcommand{\nn}{\nonumber}
\renewcommand{\d}{\mathrm{d}}
\begin{document}
\title{On a class of multidimensional integrable hierarchies and their reductions}
\author{
L.V. Bogdanov\thanks
{L.D. Landau ITP, Kosygin str. 2,
Moscow 119334, Russia, e-mail
leonid@landau.ac.ru}}
\date{}
\maketitle
\begin{abstract}
A class of multidimensional integrable hierarchies connected with commutation of 
general (unreduced) 
(N+1)-dimensional
vector
fields containing derivative over spectral variable
is considered. They are represented in the form of generating equation, as well as
in the Lax-Sato form. A dressing scheme based on nonlinear vector Riemann problem is
presented for this class. 
The hierarchies connected with 
Manakov-Santini equation and Dunajski system are considered as illustrative examples.

\end{abstract}
\section{Introduction}
Using a construction developed in \cite{BDM06,BDM07}, we describe a structure of
general integrable hierarchy connected with commutation of 
($N+1$)-dimensional vector fields
(containing a derivative over spectral variable),
which defines ($N+2$)-dimensional integrable equations. 
We consider several special cases of
the general hierarchy and its reductions, some of them defining known integrable
models. The simplest case of (1+1)-dimensional vector field 
(one space variable and a spectral variable)
corresponds to (2+1)-dimensional integrable
system introduced by Manakov and Santini \cite{MS06} (see also \cite{MS07,MS08}).
The case of (2+1)-dimensional vector fields (two space variables and a spectral variable),
after a reduction to volume-preserving vector fields, leads to
(2+2)-dimensional Dunajski integrable model \cite{Dun02}, describing anti-self-dual 
Einstein-Maxwell equation for signature (2,2)(twistor setting).

\section{General ($N+2$)-dimensional hierarchy}
To define general hierarchy, we consider $N+1$ formal
series, depending on $N$ infinite sets of
additional variables (`times')
\bea
&&
\Psi^0=\lambda+\sum_{n=1}^\infty \Psi^0_n(\mathbf{t}^1,\dots,\mathbf{t}^N)\lambda^{-n},
\label{form0}
\\&&
\Psi^k=\sum_{n=0}^\infty t^k_n (\Psi^0)^{n}+
\sum_{n=1}^\infty \Psi^k_n(\mathbf{t}^1,\dots,\mathbf{t}^N)(\Psi^0)^{-n}.
\label{formk}
\eea
where $0\leqslant k\leqslant N$, $\mathbf{t}^k=(t^k_0,\dots,t^k_n,\dots)$. 
We denote 
$\partial^k_n=\frac{\partial}{\partial t^k_n}$ and
introduce  the projectors 
$(\sum_{-\infty}^{\infty}u_n \lambda^n)_+
=\sum_{n=0}^{\infty}u_n \lambda^n$,
$(\sum_{-\infty}^{\infty}u_n \lambda^n)_-=\sum_{-\infty}^{n=-1}u_n \lambda^n$.

The general hierarchy is defined by the generating relation
\be
(J_0^{-1}\d \Psi^0\wedge \d \Psi^1\wedge \dots \wedge \d \Psi^N)_-=0,
\label{analyticity0}
\ee
where the differential includes both times and a
spectral variable,
\beaa
\d f=\sum_{k=1}^{N}\sum_{n=0}^{\infty}\partial^k_n f \d t^k_n 
+ \partial_\lambda f \d \lambda,
\eeaa
and $J_0$ is a determinant of Jacobian matrix $J_{ij}$,
$$
J_{ij}=\partial_i \Psi^j,\quad 0\leqslant i,j \leqslant N,
\quad \partial_0=\frac{\partial}{\partial \lambda},\;
\partial_k=\frac{\partial}{\partial x^k},\; \quad 1\leqslant k \leqslant N,
$$
where $x^k=t^k_0$. Equation (\ref{analyticity0}) is a generating relation for Lax-Sato 
form of the hierarchy. 
\begin{prop}
Relation
(\ref{analyticity0})
is equivalent to the set of equations
\bea
&&
\partial^k_n\mathbf{\Psi}=\sum_{i=0}^N(J^{-1}_{ki} (\Psi^0)^n)_+
{\partial_i}\mathbf{\Psi},\quad 0\leqslant n\leqslant \infty\, ,
1\leqslant k \leqslant N.
\label{genSato}
\eea
\label{formSato}
\end{prop}
The proof of this proposition is completely analogous to the 
similar proof given in \cite{BDM07} for Dunajski hierarchy. 
The proof of (\ref{analyticity0}) $\Rightarrow$
hierarchy (\ref{genSato}) is based on the following
statement.
\begin{lemma}
Given identity (\ref{analyticity0}), for arbitrary first order operator $\hat U$,
$$
\hat U\mathbf{\Psi} =\left(\sum_{k=1}^{N}\sum_i u^k_i(\lambda,\mathbf{t})\partial^k_i
+
u^0(\lambda,\mathbf{t}^1,\mathbf{t}^2)\partial_\lambda \right)\mathbf{\Psi}
$$
with `plus' coefficients ($(u^k_i)_-=u^0_-=0$), 
the condition $(\hat U\mathbf{\Psi})_+=0$
(where for $\Psi^k$, $k\neq 0$, derivatives of series (\ref{formk}) are taken with fixed $\Psi^0$)
implies that $\hat
U\mathbf{\Psi}=0$.
\label{operator}
\end{lemma}

The statement (\ref{genSato}) $\Rightarrow$ 
(\ref{analyticity0}) directly follows
from the relation
\begin{lemma}
\be
\det (\frac{\partial}{\partial\tau_i}\Psi_j)=
\det (V_{\tau_i+}^j)J_0,
\ee
where $\tau_i$, $0\leqslant i\leqslant N$,
is an arbitrary set of times of the hierarchy,
and $V_{\tau_i+}^j$ are the 
coefficients of corresponding vector fields
given by the r.h.s. of equations (\ref{genSato}),
$$
\partial_{\tau_i} \mathbf{\Psi}=\sum_{j=0}^N V_{\tau_i+}^j
{\partial_j}\mathbf{\Psi}.
$$
\end{lemma}
Similar to \cite{BDM07}, it is also possible to prove that 
Lax-Sato flows of the hierarchy (\ref{genSato}) are compatible.
\section{Manakov-Santini hierarchy}
A simplest case of general hierarchy (\ref{genSato}) with $N=1$
is connected with the system recently introduced by Manakov and Santini
\cite{MS06}. For this hierarchy we have two series
\bea
&&
\Psi^0=\lambda+\sum_{n=1}^\infty \Psi^0_n(\mathbf{t}^1,\mathbf{t}^2)\lambda^{-n},
\label{form01}
\\&&
\Psi^1=\sum_{n=0}^\infty t_n (\Psi^0)^{n}+
\sum_{n=1}^\infty \Psi^1_n(\mathbf{t})(\Psi^0)^{-n}
\label{form1}
\eea
and generating relation
\be
(J_0^{-1}\d \Psi^0\wedge \d \Psi^1)_-=0.
\label{analyticity01}
\ee
Lax-Sato equations for Manakov-Santini hierarchy read
\bea
&&
\partial_n\mathbf{\Psi}=\sum_{i=0,1}(J^{-1}_{1i} (\Psi^0)^n)_+
{\partial_i}\mathbf{\Psi},\quad 0\leqslant n\leqslant \infty\, ,
\label{genSato1}
\eea
where 
\be
J=
\begin{pmatrix}
\Psi^0_\lambda &\Psi^1_\lambda \\
\Psi^0_x&\Psi^1_x 
\end{pmatrix},
\label{J1}
\ee
and 
\beaa
&&
J^{-1}_{10}=-J_0^{-1}\Psi^0_x,\;J^{-1}_{11}=J_0^{-1}\Psi^0_\lambda,\;
\\
&&
J_0=\det J=1+\partial_x\Psi^1_1\lambda^{-1}+(\partial_x\Psi^1_2-
\Psi^0_1)\lambda^{-2}+\dots.
\eeaa
Lax-Sato equations for the first two flows are given by
\beaa
&&
\partial_y\mathbf{\Psi}=((\lambda-v_{x})\partial_x - u_{x}\partial_\lambda)\mathbf{\Psi},
\\
&&
\partial_t\mathbf{\Psi}=((\lambda^2-v_{x}\lambda+u -v_{y})\partial_x
-(u_{x}\lambda+u_{y})\partial_\lambda)\mathbf{\Psi},
\eeaa
where $u=\Psi^0_1$, $v=\Psi^1_1$. Compatibility condition for these equations
reperesents a system of equations for $u$, $v$,
\bea
u_{xt} &=& u_{yy}+(uu_x)_x+v_xu_{xy}-u_{xx}v_y,
\nn\\
v_{xt} &=& v_{yy}+uv_{xx}+v_xv_{xy}-v_{xx}v_y.
\label{MSeq}
\eea
This system was introduced by Manakov and Santini in \cite{MS06}
(see also \cite{MS07}, \cite{MS08}).
For $v=0$  the system reduces to the dKP equation
\begin{equation}
 u_{xt} = u_{yy}+(uu_x)_x.
\label{dKP-eq}
\end{equation}
Respectively, $u=0$ reduction gives an equation 
\cite{Pavlov03} (see also \cite{Dun04,MS02,MS04})
\begin{equation}
 v_{xt} = v_{yy}+v_xv_{xy} - v_{xx}v_y.
\label{Pavlov}
\end{equation}

More generally, to reduce Manakov-Santini hierarchy (\ref{genSato1})
to dispersionless KP hierarchy, one should consider a reduction $J_0=1$
(area preservation or Hamiltonian condition), then from Lax-Sato
equations (\ref{genSato1}) one easily gets Lax-Sato equations for dKP
hierarchy.

Respectively, reduction $\Psi^0=\lambda$ leads to the hierarchy
connected with equation (\ref{Pavlov}), which was considered
in \cite{MS02,MS04}.

\section{Dunajski equation hierarchy}
This  hierarchy is connected with the integrable model introduced
by Dunajski \cite{Dun02}, which generalizes the famous Plebanski 
second heavenly equation. It corresponds to general hierarchy with $N=2$
with volume-preserving reduction ($J_0=1$). To define this hierarchy,
we introduce three series
\beaa
&&
\Psi^0=\lambda+\sum_{n=1}^\infty \Psi^0_n(\mathbf{t}^1,\mathbf{t}^2)\lambda^{-n},
\\&&
\Psi^1=\sum_{n=0}^\infty t^1_n (\Psi^0)^{n}+
\sum_{n=1}^\infty \Psi^1_n(\mathbf{t}^1,\mathbf{t}^2)(\Psi^0)^{-n}
\\&&
\Psi^2=\sum_{n=0}^\infty t^2_n (\Psi^0)^{n}+
\sum_{n=1}^\infty \Psi^2_n(\mathbf{t}^1,\mathbf{t}^2)(\Psi^0)^{-n}.
\eeaa 
Generating equation for  Dunajski equation hierarchy is of the form
\be
(\d \Psi^0\wedge \d \Psi^1\wedge \d \Psi^2)_-=0
\label{analyticity0D}
\ee
(taking into account the reduction $J_0=1$).
Lax-Sato equations of the hierarchy read
\bea
\partial^1_n\mathbf{\Psi}=
+\left(
(\Psi^0)^{n}
\begin{vmatrix}
\Psi^0_\lambda & \Psi^2_\lambda\\
\Psi^0_y & \Psi^2_y
\end{vmatrix}
\right)_+\partial_x \mathbf{\Psi}-
\left(
(\Psi^0)^{n}
\begin{vmatrix}
\Psi^0_\lambda & \Psi^2_\lambda\\
\Psi^0_x & \Psi^2_x
\end{vmatrix}
\right)_+\partial_y \mathbf{\Psi}-
\nn\\
\left(
(\Psi^0)^{n}
\begin{vmatrix}
\Psi^0_x & \Psi^2_x\\
\Psi^0_y & \Psi^2_y
\end{vmatrix}
\right)_+\partial_\lambda \mathbf{\Psi}
,
\label{Dun11}
\eea
\bea
\partial^2_n\mathbf{\Psi}=
-\left(
(\Psi^0)^{n}
\begin{vmatrix}
\Psi^0_\lambda & \Psi^1_\lambda\\
\Psi^0_y & \Psi^1_y
\end{vmatrix}
\right)_+\partial_x \mathbf{\Psi}+
\left(
(\Psi^0)^{n}
\begin{vmatrix}
\Psi^0_\lambda & \Psi^1_\lambda\\
\Psi^0_x & \Psi^1_x
\end{vmatrix}
\right)_+\partial_y \mathbf{\Psi}+
\nn\\
\left(
(\Psi^0)^{n}
\begin{vmatrix}
\Psi^0_x & \Psi^1_x\\
\Psi^0_y & \Psi^1_y
\end{vmatrix}
\right)_+\partial_\lambda \mathbf{\Psi}
\label{Dun21}
\eea
(plus a condition $J_0=1$).
It is easy to check that for $\Psi^0=\lambda$ Dunajski equation hierarchy reduces
to heavenly equation hierarchy
\cite{Takasaki,Takasaki1}, while for $\Psi^2=y$ it reduces to dispersionless
KP hierarchy. The first two flows of the hierarchy (\ref{Dun11}), (\ref{Dun21})
read
\bea
&&\p_1^1\mathbf{\Psi}=(\lambda\p_x
-v_{x}\p_x-u_{x}\p_y-f_{x}\p_{\lambda})\mathbf{\Psi},
\label{fl01}
\\
&&\p_1^2\mathbf{\Psi}=(\lambda\p_y
-v_{y}\p_x
-u_{y}\p_y -f_{y}\p_{\lambda})\mathbf{\Psi},
\label{fl02}
\eea
here
$$
u=\Psi^2_1,\quad v=\Psi^1_1, \quad f=\Psi^0_1,
$$
condition $J_0=1$ implies that $u_y+v_x=0$, and we can introduce
a potential $\Th$, $v=\Th_y$, $u=-\Th_x$.
After the identification
$z=-t^1_1$, $w=t^2_1$, we get Lax pair for Dunajski system
\beaa
&&
\Th_{wx}+\Th_{zy}+\Th_{xx}\Th_{yy}-\Th_{xy}^2=f,
\\
&&
f_{xw}+f_{yz}+
\Th_{yy}f_{xx}+\Th_{xx}f_{yy}-2\Th_{xy}f_{xy}=0.
\eeaa
The second flows can be written in the form
\beaa
&&\p_2^1\mathbf{\Psi}=
(\lambda \p_1^1+f \p_x
-(\p_1^1 v)\p_x-(\p_1^1 u)\p_y-(\p_1^1f)\p_{\lambda}
)
\mathbf{\Psi},
\label{fl11}
\\
&&\p_2^2\mathbf{\Psi}=
(\lambda \p_1^2+f \p_y
-(\p_1^2 v)\p_x-(\p_1^2 u)\p_y-(\p_1^2f)\p_{\lambda}
)
\mathbf{\Psi}.
\label{fl12}
\eeaa
To write down vector fields more explicitly,
one should use first flows (\ref{fl01}), (\ref{fl02}). Commutation relations
for any pair of the flows give an equation for $\Th$, $f$ 
(with different set of times).

\section{Dressing scheme}
Let us consider nonlinear vector  Riemann problem of the form
\bea
\mathbf{\Psi}_+= \mathbf{F}(\mathbf{\Psi}_-),
\label{Riemann}
\eea
where $\mathbf{\Psi}_+$, $\mathbf{\Psi}_-$ denote the boundary values
of the (N+1)-component vector function on the sides of some
oriented curve $\gamma$ in the complex plane of the variable
$\lambda$. The problem is to find the function analytic
outside the curve with some fixed behavior at infinity (normalization) 
which satisfies
(\ref{Riemann}). The problem of this type was used by Takasaki 
\cite{Takasaki0,Takasaki,Takasaki1}, who stressed its connection to Penrose nonlinear
graviton construction.

The  problem (\ref{Riemann}) is connected with a class of integrable 
equations,
which can be represented as a commutation relation for 
vector fields containing
a derivative on the spectral variable.
We give a sketch of the dressing scheme
corresponding to the general (N+2)-dimensional integrable hierarchy 
(\ref{analyticity0}), (\ref{genSato}). First, we chose $\gamma$ as a unit circle.
To get solutions of the general hierarchy using the problem (\ref{Riemann}),
one should find solution to this problem with the singulariries at $\lambda=\infty$
defined by the series (\ref{form0}), (\ref{formk}) 
(for simplicity we suggest that only a final number of times is not equal to zero).
If the determinant $J_0$ for this solution is not equal to zero,
we come to the conclusion that the form
$$
\Omega=J_0^{-1}\d \Psi^0\wedge \d \Psi^1\wedge \dots \wedge \d \Psi^N
$$
is analytic in the complex plane,
and so it satisfies the generating identity (\ref{analyticity0}).
Then the solution of the Riemann problemm (\ref{Riemann}) satisfies
Lax-Sato equations (\ref{genSato}) and provides a solution for the general
(N+2)-dimensional hierarchy. 

Due to the fact that singularities
of solution of the Riemann problem are defined by the series (\ref{form0}), (\ref{formk})
using the solution itself, solution of the hierarchy is usually defined in 
terms of implicit functions.

\section*{Acknowledgments}
The author was partially supported by Russian Foundation for
Basic Research under grants no. 08-01-90104,  
07-01-00446 and 06-01-92053.

\end{document}